\documentclass[11pt]{article}
\usepackage{amsmath,amssymb,bm}
\usepackage{graphics}
\usepackage{graphicx}
\usepackage{amssymb}
\usepackage{epstopdf}

\usepackage{mystyle}
\usepackage{cite,./mcite}
\usepackage{authblk}

\usepackage{subfigure}

\newcommand{\tvec}[1]{\boldsymbol{#1}}

\newcommand{\ms}{\mskip 1.5mu}

\def\lsim{\mathrel{\rlap{\lower4pt\hbox{\hskip1pt$\sim$}}
    \raise1pt\hbox{$<$}}}                
\def\gsim{\mathrel{\rlap{\lower4pt\hbox{\hskip1pt$\sim$}}
    \raise1pt\hbox{$>$}}}                

\title{
\begin{flushright}
{\small DESY 16--048}
\end{flushright}
Double parton scattering in the ultraviolet: \\
  addressing the double counting problem\thanks{To appear in the
  proeceedings of MPI@LHC 2015, Trieste, Italy, 23--27 November 2015}}

\author[1]{Markus Diehl}
\author[2]{Jonathan R.\ Gaunt}
{\tiny
\affil[1]{Deutsches Elektronen-Synchroton DESY, 22603 Hamburg, Germany}
\affil[2]{Nikhef Theory Group and VU University Amsterdam, De Boelelaan
  1081, 1081 HV Amsterdam, The Netherlands}
}

\date{\parbox{0.9\textwidth}{\small \textbf{Abstract:} In proton-proton
collisions there is a smooth transition between the regime of double
parton scattering, initiated by two pairs of partons at a large relative
distance, and the regime where a single parton splits into a parton pair
in one or both protons.  We present a scheme for computing both
contributions in a consistent and practicable way.}}

\begin{document}

\maketitle

\section{Ultraviolet behaviour of double parton scattering}

\begin{figure}
\begin{center}
\subfigure[]{\includegraphics[height=4.5em]{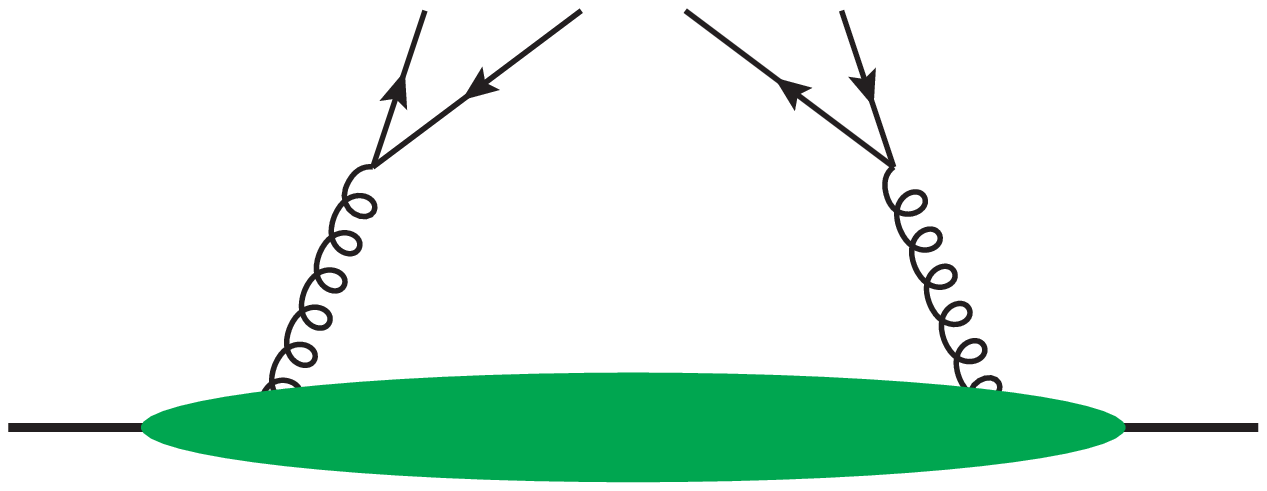}}
\hspace{1em}
\subfigure[]{\includegraphics[height=8em]{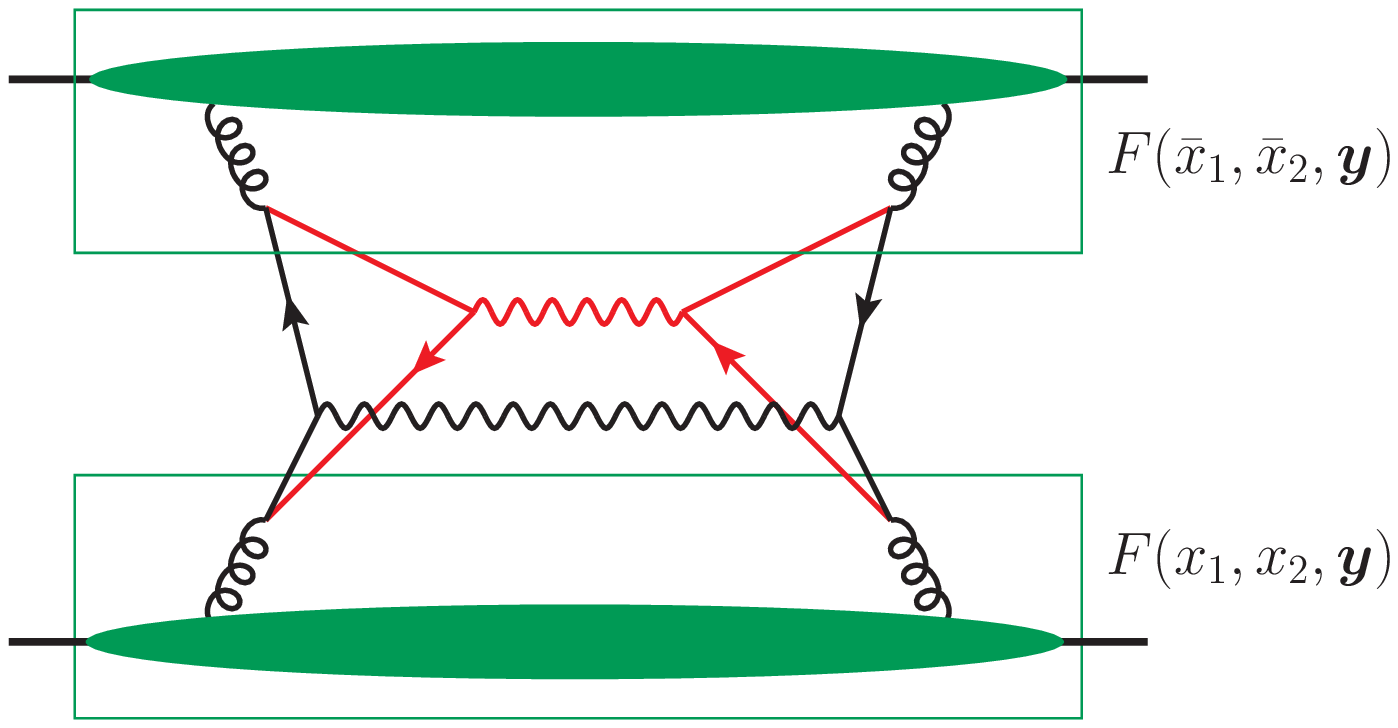}}
\hspace{1em}
\subfigure[]{\includegraphics[height=8em]{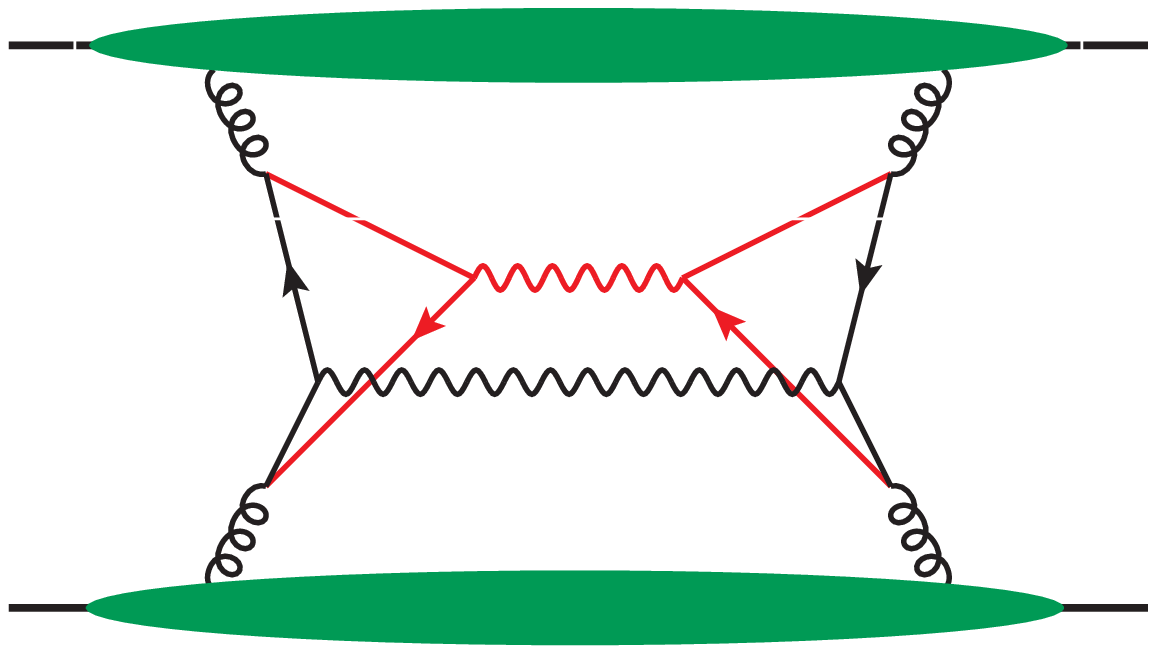}}
\caption{\label{fig:split} (a) Perturbative splitting contribution to a
  DPD.  (b) Contribution of double perturbative splitting to DPS, also
  called ``1 vs 1'' graph.  (c) Single hard scattering contribution.}
\end{center}
\end{figure}

The familiar factorisation formula for double parton scattering (DPS)
reads
\begin{align}
  \label{dps-Xsect}
 \frac{d\sigma_{\text{DPS}}}{dx_1\ms d\bar{x}_1\,
  dx_2\ms d\bar{x}_2}
&= \frac{1}{C}\; \hat{\sigma}_1\ms \hat{\sigma}_2
\int d^2\tvec{y}\;
F(x_1, x_2, \tvec{y}) \, F(\bar{x}_1, \bar{x}_2, \tvec{y}) \,,
\end{align}
where $C$ is a combinatorial factor, $\hat{\sigma}_{1,2}$ is the cross
section for the first or second hard-scattering subprocess, and
$F(x_1,x_2, \tvec{y})$ is a double parton distribution (DPD).  $\tvec{y}$
denotes the transverse distance between the two partons.  A field
theoretical definition of $F(x_1,x_2, \tvec{y})$ is naturally given by the
matrix element between proton states of two twist-two operators at
relative transverse distance $\tvec{y}$.  As explained in
\cite{Diehl:2011yj}, the leading behaviour of DPDs at small $\tvec{y}$ is
controlled by the splitting of one parton into two, shown in
figure~\ref{fig:split}a.  The corresponding expression reads
\begin{align}
  \label{split-dpd}
F(x_1,x_2, \tvec{y}) &= \frac{1}{\tvec{y}^2}\, \frac{\alpha_s}{2\pi^2}\,
\frac{f(x_1+x_2)}{x_1+x_2}\, T\biggl( \frac{x_1}{x_1+x_2} \biggr) & &
\text{for small $\tvec{y}$} \,.
\end{align}
For simplicity we dropped labels for the different parton species and
polarisations, as we already did in \eqref{dps-Xsect}.

Inserting the short-distance limit \eqref{split-dpd} in the cross-section
formula \eqref{dps-Xsect} reveals an immediate problem: the integration
over $\tvec{y}$ diverges strongly in the ultraviolet.  In fact, the
approximations that lead to \eqref{dps-Xsect} are not valid when
$\tvec{y}$ becomes too small (compared with the inverse of the large
momentum scale $Q$ of the hard scattering).  This unphysical ultraviolet
divergence signals another problem, namely one of double counting: the
graph in figure~\ref{fig:split}b shows a contribution to double parton
scattering, with perturbative splitting in each DPD.  Drawn as in
figure~\ref{fig:split}c, the same graph gives however a contribution to
single parton scattering (SPS) at higher loop order.  For multi-jet
production this problem was already pointed out in \cite{Cacciari:2009dp}.


\section{A consistent scheme}

The following scheme provides a consistent treatment of single and double
scattering contributions to a given process, and it removes the
ultraviolet divergence in the naive double scattering formula just
discussed.  We regulate the DPS cross section \eqref{dps-Xsect} by
inserting a function under the integral over DPDs,
\begin{align}
  \label{dps-reg}
\int d^2\tvec{y}\; \bigl[ \Phi(\nu y) \bigr]{}^2\,
  F(x_1, x_2, \tvec{y}) \, F(\bar{x}_1, \bar{x}_2, \tvec{y}) \,,
\end{align}
which is chosen such that $\Phi(u) \to 0$ for $u\to 0$ and $\Phi(u) \to 1$
for $u \gg 1$.  (We take the square of $\Phi$ in order to have a closer
connection to the case discussed in section~\ref{sec:tmds}.)  This removes
contributions with distances $y = |\tvec{y}|$ below $1/\nu$ from what is
\emph{defined} to be double parton scattering.  An appropriate choice for
this cutoff scale is $\nu \sim Q$.  Double and single parton scattering
are then combined as
\begin{align}
  \label{full-Xsect}
\sigma_{\text{DPS}} - \sigma_{\text{sub}} + \sigma_{\text{SPS}} \,,
\end{align}
where $\sigma_{\text{DPS}}$ is the regulated DPS cross section and
$\sigma_{\text{SPS}}$ the SPS cross section computed in the usual way
(given by figure~\ref{fig:split}c and its crossed variants in our
example).  The subtraction term $\sigma_{\text{sub}}$ is given by the DPS
cross section with both DPDs replaced by the splitting expression
\eqref{split-dpd}, computed at fixed order in perturbation theory and used
at all $\tvec{y}$.  Note that at any order in $\alpha_s$, the computation
of $\sigma_{\text{sub}}$ is technically much simpler than the one of
$\sigma_{\text{SPS}}$.

Let us see how this construction solves the double counting problem.  We
work differentially in $\tvec{y}$, which is Fourier conjugate to a
specific transverse momentum variable as specified in \cite{Diehl:2011yj}
and can thus be given an unambiguous meaning, not only in the DPS cross
section but also in the box graph of figure~\ref{fig:split}c and the
associated term $\sigma_{\text{SPS}}$.  For $y \lsim 1/Q$ one has
$\sigma_{\text{DPS}} \approx \sigma_{\text{sub}}$ because the perturbative
approximation \eqref{split-dpd} of the DPD works well in that region.  The
dependence on the cutoff function $\Phi(\nu y)$ then cancels between
$\sigma_{\text{DPS}}$ and $\sigma_{\text{sub}}$, and one is left with
$\sigma \approx \sigma_{\text{SPS}}$.  For $y \gg 1/Q$ one has
$\sigma_{\text{sub}} \approx \sigma_{\text{SPS}}$, because in that region
the box graph can be approximated just as is done in the DPS formula.
One is thus left with $\sigma \approx \sigma_{\text{DPS}}$ at large $y$,
and the cutoff function $\Phi(y \nu) \approx 1$ does not have any effect
there.  The construction just explained is a special case of the general
subtraction formalism discussed in chapter 10 of \cite{Collins:2011zzd},
and it works order by order in perturbation theory.


\section{Splitting and intrinsic contributions to DPDs}

At small $\tvec{y}$ a DPD -- defined as a hadronic matrix element as
already mentioned -- contains not only the perturbative splitting
contribution described by \eqref{split-dpd} but also an ``intrinsic'' part
in which the two partons do not originate from one and the same ``parent''
parton.  We emphasise that our scheme does not need to distinguish these
``splitting'' and ``intrinsic'' contributions when setting up the
factorisation formula for the cross section.  In fact, we do not know how
such a separation could be realised in a field theoretic definition valid
at all $\tvec{y}$.  It is only when writing down a parameterisation of
$F(x_1,x_2, \tvec{y})$ that has the small-$y$ limit predicted by QCD that
we separate the DPD into splitting and intrinsic pieces.

\begin{figure}
\begin{center}
\subfigure[]{\includegraphics[height=8em]{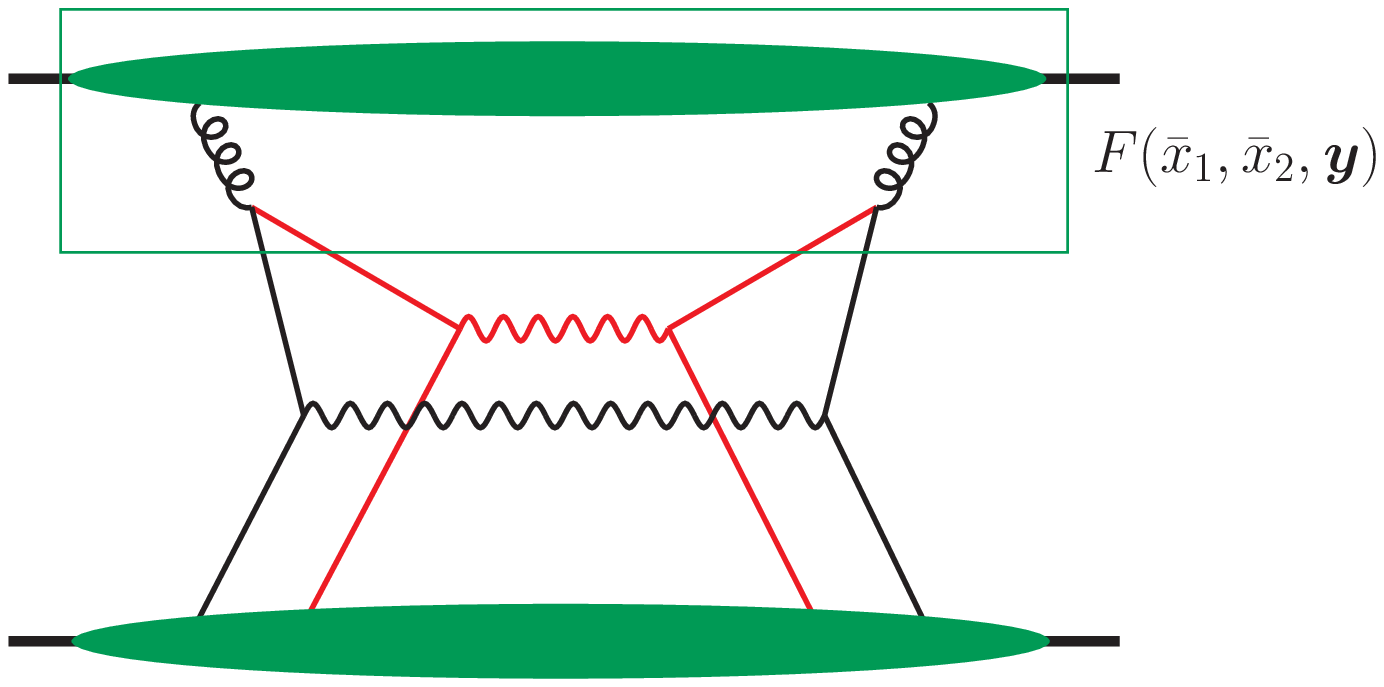}}
\hspace{1em}
\subfigure[]{\includegraphics[height=8em]{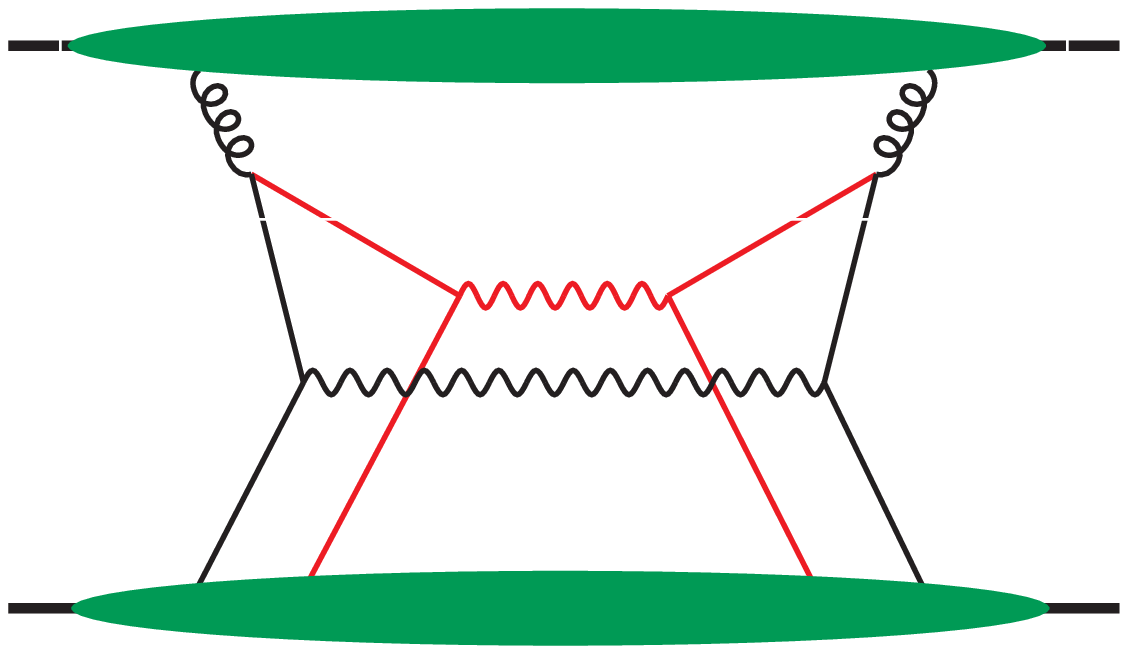}}
\caption{\label{fig:1vs2} (a) Contribution of single perturbative
  splitting to DPS, also called ``1 vs 2'' graph.  (b) Graph with a
  twist-two distribution for one proton and a twist-four distribution for
  the other.}
\end{center}
\end{figure}

If we consider the DPS cross section formula at small $y$ and take the
splitting contribution for only one of the two protons, then we obtain the
``1 vs 2'' contribution depicted in figure~\ref{fig:1vs2}a, which has been
discussed in detail in \cite{Blok:2011bu,Blok:2013bpa},
\cite{Ryskin:2011kk,Ryskin:2012qx} and \cite{Gaunt:2012dd}.  The
corresponding integral in the cross section goes like $d^2\tvec{y}
/\tvec{y}^2$ and thus still diverges at small $y$ if treated naively.  In
our regulated DPS integral~\eqref{dps-reg}, it gives a finite contribution
with a logarithmic dependence on the cutoff scale $\nu$.

Just as the 1 vs 1 contribution of figure~\ref{fig:split}b corresponds to
the SPS graph ~\ref{fig:split}c, the 1 vs 2 contribution of
figure~\ref{fig:1vs2}a corresponds to a contribution with a twist-two
distribution (i.e.\ a usual parton density) for one proton and a
twist-four distribution for the other proton, shown in
figure~\ref{fig:1vs2}b.  The complete cross section is then obtained as
\begin{align}
  \label{all-Xsect}
\sigma &= \sigma_{\text{DPS}}
   - \sigma_{\text{sub (1vs1)}} + \sigma_{\text{SPS}}
   - \sigma_{\text{sub (1vs2)}} + \sigma_{\text{tw2 $\times$ tw4}} \,.
\end{align}
The DPS term contains the full DPDs and thus generates 1 vs 1, 1 vs 2 and
the usual 2 vs 2 contributions.  The terms $\sigma_{\text{sub (1vs1)}}$
and $\sigma_{\text{SPS}}$ were discussed in the previous section.  The
term $\sigma_{\text{tw2 $\times$ tw4}}$ corresponds to
figure~\ref{fig:1vs2}b, and the associated subtraction term
$\sigma_{\text{sub (1vs2)}}$ is obtained from the DPS formula by replacing
one DPD with its perturbative splitting approximation \eqref{split-dpd}
and the other DPD with a twist-four distribution.

Since very little is known about parton distributions of twist four,
including $\sigma_{\text{tw2 $\times$ tw4}} - \sigma_{\text{sub (1vs2)}}$
in the cross section is a challenge for phenomenology.  One can however
show that with the choice $\nu \sim Q$ this combination is subleading in
logarithms $\log(Q/\Lambda)$ compared to the 1 vs 2 part of
$\sigma_{\text{DPS}}$ and can hence be dropped at leading logarithmic
accuracy.


\section{DGLAP logarithms}

\begin{figure}
\begin{center}
\subfigure[]{\includegraphics[height=8em]{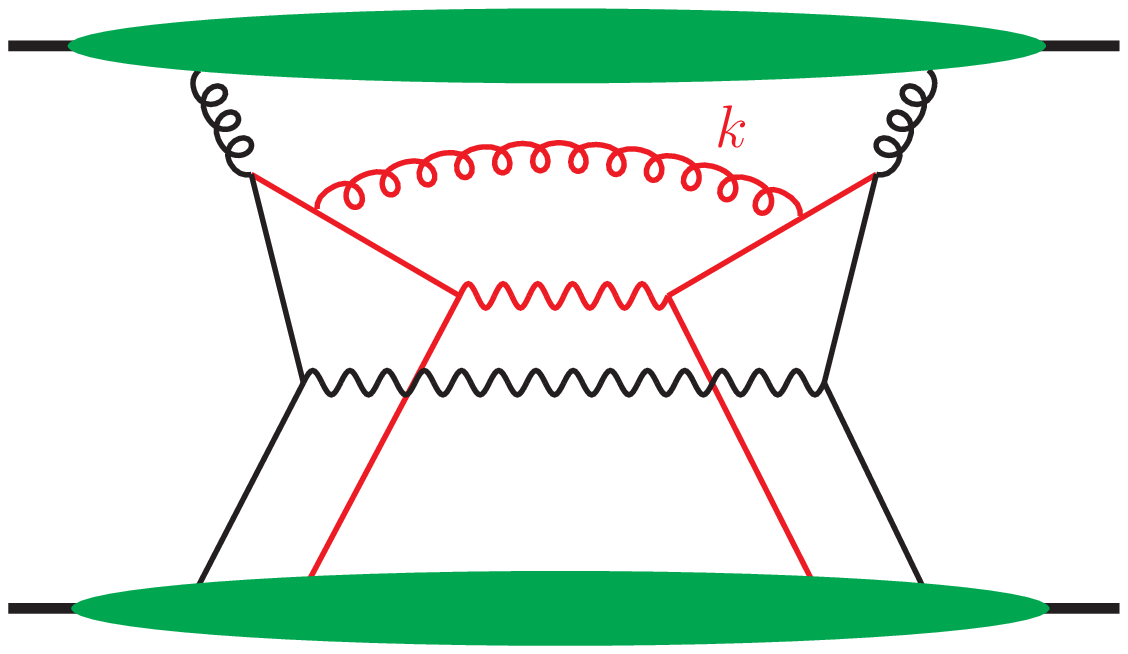}}
\hspace{1em}
\subfigure[]{\includegraphics[height=8em]{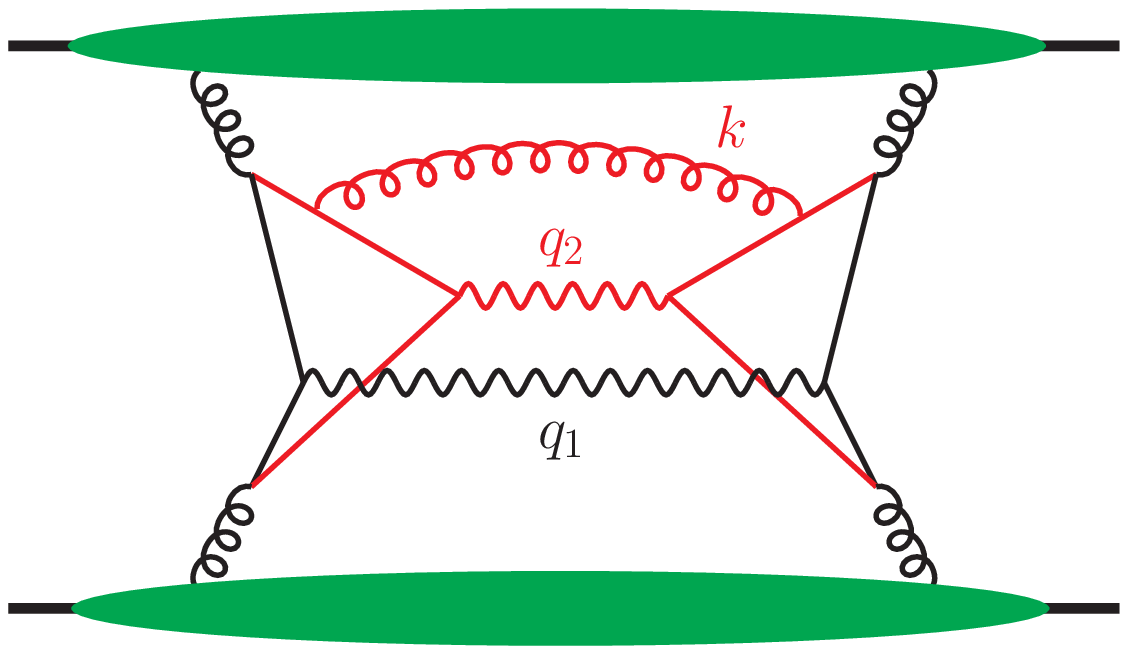}}
\caption{\label{fig:dglap} Graphs with additional gluon emission that give
  rise to DGLAP logarithms in strongly ordered kinematics, as explained in
  the text.}
\end{center}
\end{figure}

As discussed in \cite{Diehl:2011yj}, the DPDs $F(x_1, x_2, \tvec{y})$ are
subject to homogeneous DGLAP evolution, with one DGLAP kernel for the
parton with momentum fraction $x_1$ and another for the parton with
momentum fraction $x_2$.  One can show that the evolved distributions in
the DPS cross section correctly resum large DGLAP logarithms in
higher-order graphs.  An example is the 1 vs 2 graph in
figure~\ref{fig:dglap}a, which builds up a logarithm $\log^2(Q/\Lambda)$
in the region $\Lambda \ll |\tvec{k}| \ll Q$, compared with the single
$\log(Q/\Lambda)$ of the graph without gluon emission.  Similarly, the
higher-order SPS graph in figure ~\ref{fig:dglap}b builds up a logarithm
$\log(Q_2/Q_1)$ in the region $Q_1 \ll |\tvec{k}| \ll Q_2$ if the scales
$Q_1$ and $Q_2$ of the two hard scatters are strongly ordered among
themselves.  This type of logarithm is readily included in the cross
section by taking separate renormalisation scales $\mu_{1,2} \sim Q_{1,2}$
for the two partons in the DPDs.

\section{Extension to measured transverse momenta}
\label{sec:tmds}

So far we have discussed DPS and SPS in collinear factorisation, where the
net transverse momentum $\tvec{q}_{1}$ and $\tvec{q}_{2}$ of the particles
produced by each hard scatter is integrated over.  As shown in
\cite{Diehl:2011yj}, DPS can also be formulated for small measured
$\tvec{q}_{1}$ and $\tvec{q}_{2}$ by generalising the corresponding
formalism for SPS (which is e.g.\ documented in chapter~13 of
\cite{Collins:2011zzd}).  Our scheme is readily extended to this case.
The DPS cross section then involves a regularised integral
\begin{align}
  \label{tmd-reg}
\int d^2\tvec{y}\, d^2\tvec{z}_1\, d^2\tvec{z}_2\;
  e^{- i \tvec{q}_1 \tvec{z}_1 - i \tvec{q}_2 \tvec{z}_2}\,
  \Phi(\nu y_+)\, \Phi(\nu y_-)\,
  F(x_1,x_2, \tvec{z}_1,\tvec{z}_2, \tvec{y}) \,
  F(\bar{x}_1,\bar{x}_2, \tvec{z}_1,\tvec{z}_2, \tvec{y}) \,,
\end{align}
where $F(x_1,x_2, \tvec{z}_1,\tvec{z}_2, \tvec{y})$ is a
transverse-momentum dependent DPD transformed to impact parameter space.
The perturbative splitting mechanism renders these distributions singular
at the points $y_\pm = \bigl| \tvec{y} \pm \tfrac{1}{2} (\tvec{z}_1 -
\tvec{z}_2) \bigr|$, as seen in section~5.2 of \cite{Diehl:2011yj}, and
the function $\Phi$ regulates the logarithmic divergences that appear in
the naive DPS formula.


\section*{Acknowledgements}

J.G.\ acknowledges financial support from the European Community under the
Ideas program QWORK (contract 320389).

\bibliographystyle{MPI2015} 
\bibliography{diehl}

\end{document}